\documentclass[final,english]{bullsrsl}[2022/06/15]

\usepackage[latin1]{inputenc}
\usepackage[T1]{fontenc}

\usepackage{natbib} 
\usepackage{graphicx}

\usepackage{hyperref}
\hypersetup{
    colorlinks = true,
    allcolors = {blue}
}

\begin{document}
\title{Chemodynamic analysis of four $r$-process enhanced stars observed with GTC}

\author[affil={1,2}, corresponding]{}{Pallavi Saraf}
\author[affil={1}]{Thirupathi Sivarani}{}

\affiliation[1]{Indian Institute of Astrophysics, Koramangala 2nd Block, Bangalore, 560034, India}
\affiliation[2]{Pondicherry University, R.V. Nagar, Kalapet, 605014, Puducherry, India}
\correspondance{pallavi.saraf@iiap.res.in}
\date{15th April 2023}
\maketitle


%

\begin{abstract}
Here, we delineate a comprehensive abundance analysis of four $r$-process enhanced metal-poor stars observed with HORuS spectrograph on a 10-m class telescope, GTC. The high signal-to-noise ratio at $R \approx 25000$ spectral resolution allowed us to detect 16 light and 20 neutron-capture elements along with Th in two stars. Four of our program stars show signatures of mixing in their atmosphere. Through detailed abundance analysis of four $r$-process enhanced stars together with already identified $r$-process-rich stars in literature, we probe the production sites of neutron-capture elements. The [Zr/Eu] ratio as a function of metallicity shows the evidence of multiple channels for the production of $r$-process. Thorium to first and second $r$-process peak elements ratios also support the similar non-universality of neutron-capture elements. An increased sample of $r$-process enhanced stars will enable us understand different formation channels of neutron capture elements. Using the kinematic analysis, we found the clues of accretion for two of our program stars.
\end{abstract}


\section{Introduction}
The astrophysical conditions and sites for the production of elements beyond the iron group is one of the poorly understood questions in the field of stellar nucleosynthesis. Elements up to the iron group are processed through fusion reaction. The formation of heavier elements beyond Z=30 is prevented by nuclear repulsion resulting from the lower binding energy per nucleon. In this situation, the successive neutron-capture produces heavier nuclei above the iron group by overcoming the barrier of nuclear repulsion \citep{Burbidge.etal.1957, Cameron.1957}. Depending on the neutron capturing rate, neutron-capture processes are categorized into slow neutron-capture process (or in short 'the $s$-process') and rapid neutron-capture process (or in short 'the $r$-process'). The $s$-process, for which the neutron-capture timescale is much longer than the beta-decay timescale, requires neutron density of 10$^{8}$ cm$^{-3}$. However, the $r$-process, for which the neutron-capture timescale is much smaller than the beta-decay timescale, needs a neutron density greater than 10$^{22}$ cm$^{-3}$. According to theoretical calculations, during the $r$-process, seed nuclei can capture approximately 100 neutrons per second. In contrast, during the $s$-process, the seed nuclei capture only one neutron every 1000 years \citep{Meyer.1994}.

Our understanding of the astrophysical sites and the conditions that give rise to the $s$-process has significantly advanced. However, we are still unclear about the astrophysical environment for $r$-process. The $s$-process primarily occurs in low- to intermediate-mass stars during their advanced evolutionary stages, such as during the asymptotic giant branch (AGB) phase \citep{Herwig.2005, Doherty.etal.2015}. These stars possess the necessary conditions for the $s$-process to take place, including the appropriate temperature, density, and neutron flux. Numerous theoretical models have been put forward to account for the observed chemical evolution of the $r$-process in the Galaxy, including scenarios involving neutron stars mergers (NSM) \citep{Lattimer.Schramm.1974, Rosswog.etal.2014}, magneto-rotational supernovae \citep{Woosley.1993, Siegel.Metzger.2018}, and collapsars \citep{Siegel.etal.2019}. However, none of these proposed models are sufficiently well-constrained. The lack of definitive evidence or conclusive data hampers our ability to pinpoint the exact astrophysical environments where the $r$-process occurs.

The recent discoveries of NSM by the LIGO and Virgo gravitational wave observatories have provided compelling evidence supporting NSM as potential sites for the $r$-process \citep{Arcavi.etal.2017, Tanvir.etal.2017}. The electromagnetic counterpart of NSM (also known as kilonova) supplied the evidence of Sr production \citep{Watson.etal.2019}. This detection strengthens the connection between NSM events and the rapid neutron-capture process, further solidifying NSM as a significant astrophysical site for the creation of heavy elements. Apart from this, the detection of a large number of highly $r$-process enhanced stars in ultra-faint dwarf galaxy Reticulum II (Ret II) also supports the NSM as a primary site for $r$-process nucleosynthesis \citep{Ji.etal.2016}. The notable loophole of NSM is its delayed merge time which poses a challenge when explaining the presence of $r$-process enhanced stars during the early stages of Galactic Chemical Evolution.

The surface elemental abundances of metal-poor stars ([Fe/H] < -1.0) hold significant value when studying early $r$-process enrichment. These stars serve as exceptional sources of information as they retain the imprints of their natal clouds that were enriched with $r$-process material \citep{sneden.etal.2008}. The natal clouds from which these metal-poor stars are formed contain the remnants of explosive astrophysical events where the $r$-process played a vital role in synthesizing heavy elements. By examining the elemental abundances of these stars, particularly the presence of $r$-process elements, we can gain valuable insights into the chemical conditions prevailing during the early stages of the Universe.

According to the enhancement level of $r$-process yield, the stellar astrophysicists have classified the $r$-process stars into three sub-categories based on the $r$-process element Europium, as it is easily detectable in the optical spectra. These are $r$-I (+0.3 $\leq$ [Eu/Fe] $\leq$ +0.7 and [Ba/Eu] $<$ 0.0), $r$-II ([Eu/Fe] $>$ +0.7 and [Ba/Eu] $<$ 0.0) and limited-$r$ ([Eu/Fe] $<$ 0.3, [Sr/Ba] $>$ 0.5, and [Sr/Eu] $>$ 0.0) according to recent classification by \citep{Holmbeck.etal.2018}.

In this study, we perform a detailed abundance analysis of four $r$-process enhanced stars. We have also collected chemical abundances of $r$-process enhanced stars available in the literature. The main objective of this study is to investigate the production $r$-process elements and to find a correlation among $r$-I, $r$-II, and limited-$r$ stars.





We have arranged this work in the following manner, Section~\ref{sec:data} talks about the data used for this study. In Section~\ref{sec:stellar_param}, we have discussed our methodology for stellar parameter estimation. Section~\ref{sec:results} presents results of this study. Finally, Section~\ref{sec:conclusions} summarises our findings.

\section{Observation}
\label{sec:data}
For this work, we choose four metal-poor stars that are relatively bright with a V-band magnitude less than 12. Initially, we identified these stars from the SDSS-III MARVELS radial velocity survey. To obtain high-resolution ($R \sim 25000$) and high signal-to-noise ratio (SNR) spectra, we observed these objects using the High Optical Resolution Spectrograph (HORuS) installed on the Gran Telescopio Canarias (GTC) \citep{HORUS.2020}. GTC is a 10-m class reflecting telescope situated at Roque de Los Muchachos Observatory on the island of La Palma in the Canaries, Spain.

To process the spectroscopic data of our program stars, we employed a well-established and widely accepted procedure. The entire data reduction process was executed using the advanced tools available within the Image Reduction and Analysis Facility (IRAF) developed by the National Optical Astronomy Observatory (NOAO). Specifically, we utilized the IMRED, CCDRED, and ECHELLE packages of IRAF to carry out the necessary operations for our data reduction.

Subsequently, we determined the radial velocities (RVs) of the stars in our study by comparing the observed spectra to template synthetic spectra with similar metallicities and temperatures. To perform this task, we used an IDL routine CRSCOR. The resulting estimated radial velocity was then corrected for the motion of the earth around the Sun using the RVCORRECT task within IRAF. For further analysis, we accounted for the Doppler shift in the observed spectra due to radial velocity.

\section{Estimation of Stellar parameter and Abundance}
\label{sec:stellar_param}
Throughout this work, we make use of  ATLAS9 \citep{ATLAS.1993, ATLAS.2003} for atmospheric model generation and TURBOSPECTRUM \citep{Turbospectrum.1998, Turbospectrum.2012} for abundance estimation. The abundances of Fe absorption features are used for the estimation of spectroscopic parameters.

\textbf{Effective Temperature:}
To obtain an initial estimation of the temperature, we utilized various photometric colors such as V-K, J-H, and J-K obtained from existing literature. These colors served as indicators for determining the temperature of the stars under investigation. Additionally, for temperature estimation based on Spectral Energy Distribution (SED), we employed an online tool called VOSA (Virtual Observatory Sed Analyzer). To estimate the stellar parameters through the spectroscopic method, we identified specific absorption lines of Fe I and Fe II in the spectra. The spectroscopic temperature was determined by assuming excitation equilibrium between the neutral iron lines, meaning that the abundance of the Fe I line remained consistent regardless of the lower excitation potential (LEP).

\textbf{Surface Gravity:} 
To estimate the surface gravity ($\log g$) of the stars, we employed the assumption of ionization balance for a specific element. This assumption states that the abundance of an element, determined either from its neutral state (Fe-I) or ionized state (Fe-II), remains consistent. Furthermore, we utilized the fitting of magnesium triplet regions as an additional method to calculate the surface gravity of the stars under investigation. To ensure consistency, we plotted our program stars in a color-magnitude diagram (CMD). All these methods yield consistent results and exhibit good agreement with one another.

\textbf{Micro-turbulence:}
In order to determine the micro-turbulence velocity, we enforced the condition that the abundance of the neutral iron line (Fe I) remains unaffected of its reduced equivalent width. As an additional verification step, we computed the micro-turbulence velocity using a methodology outlined in previous studies such as \cite{Sahin.Lambert.2009}, \cite{Reddy.etal.2012}, and \cite{Molina.etal.2014}. This approach utilizes the dispersion in the element's abundance. The micro-turbulence velocity of the atmosphere is identified as the value at which the dispersion is minimized. Table~\ref{tab:Adopted_stellar_parameters} lists the stellar parameters of our program stars.

\begin{table}[]
	\centering
	\caption{Final adopted stellar parameters for our program stars.}
	\label{tab:Adopted_stellar_parameters}
	\begin{tabular}{lcccc} 
		\hline
		Star name & $T_{\rm eff}$ & $\log g$ & [Fe/H] & $\xi$\\
		 & (K) &  &  & (km/s) \\
		\hline
		TYC 3431-689-1 & 4850 & 1.4 & -2.05 & 1.80 \\
		HD 263815 & 4700 & 1.3 & -2.22 & 1.85\\
		TYC 1191-918-1 & 4650 & 1.1 & -1.90 & 1.95\\
		TYC 1716-1548-1 & 4500 & 1.2 & -2.15 & 1.95\\
		\hline
	\end{tabular}
\end{table}

\textbf{Abundance Analysis:}
We have measured the equivalent widths of the lines of neutral and ionized atoms of most of the elements. Abundances of the elements are determined from the measured equivalent width of the lines using a recent version of Turbospectrum 19 code and the adopted model atmosphere. We have also performed spectrum synthesis calculations of all the elements present in the spectra. Due to the good signal-to-noise ratio of our program stars, we could calculate 16 light elements namely CH, CN, Mg, Si, Ca, Ti, Na, Al, Sc, V, Cr, Mn, Fe, Co, Ni, Cu, Zn along with 20 neutron-capture elements namely Sr, Y, Zr, Ba, La, Ce, Pr, Nd, Sm, Eu, Gd, Tb, Dy, Ho, Er, Tm, Lu, Hf, Os, Th among four of our program stars (\cite{Saraf.etal.2023}. All the light elements are consistent with normal halo stars, and heavy elements are consistent with the solar r-process pattern. 

\section{Results and discussion}
\label{sec:results}

\subsection{Chemical peculiarity of program stars}
The elemental abundances of C, Fe, Sr, Ba, and Eu are used to classify stars into different chemical sub-classes. We could obtain the abundances of these key elements in four of our program stars. All of our program stars show [C/Fe] ratio lower than 0.7 dex, which makes them carbon-normal stars. Also, we found that [Ba/Eu] elemental ratio is lower than 0.0 dex for all four stars suggesting they are enhanced in r-process elements. We found that HD 263815 and TYC 1191-918-1 fall in the $r$-I category with [Eu/Fe] in the range from 0.3 to 0.7 dex, whereas TYC 3431-689-1 and TYC 1716-1548-1 lies in $r$-II category with [Eu/Fe] greater than 0.7 dex.

\subsection{Evolution of elements upto Zn}
\begin{figure}
    \centering
    \includegraphics[width=0.6\textwidth]{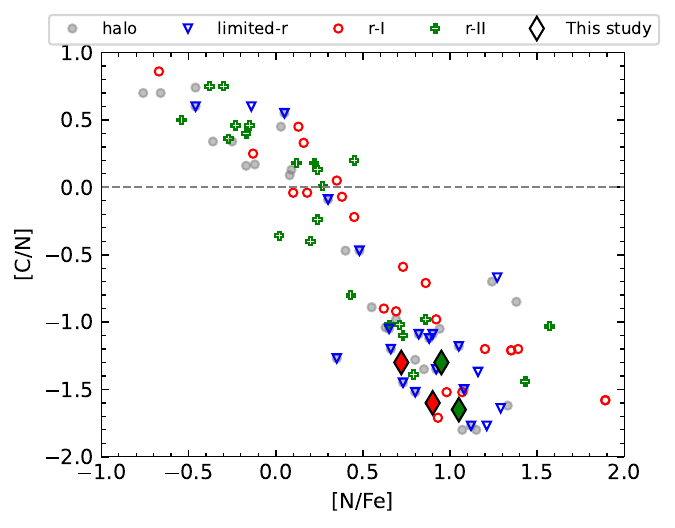}
    \caption{Carbon to nitrogen ratio as a function of nitrogen to iron ratio. Red open circles show $r$-I stars, green open crosses display $r$-II stars, blue down triangles represent limited-$r$ stars, and gray filled circles are normal halo stars. Big red and green faced diamonds are the stars from this study. The figure shows that four of our program stars are falling in the mixed region defined by [C/N] $<$ -1.0 and [N/Fe] $>$ 0.5 dex.}
    \label{fig:N_Fe__C_N}
\end{figure}
We found that four of our program stars show a low [C/Fe] ratio in the range from -0.70 to -0.35 dex. It is due to their current evolutionary stage. After correcting for their evolution, the [C/Fe] ratio ranges from -0.05 to +0.27 dex, and thus our program stars move in the region of the other $r$-process enhanced stars. To further investigate it, we have shown the [C/N] ratio as a function of [N/Fe] for our program stars in figure~\ref{fig:N_Fe__C_N}. Four of our program stars show [C/N] $<$ -1.0 and [N/Fe] $<$ -0.5 dex. It confirms that these objects have undergone the mixing process.

The elemental abundances of $\alpha$ elements (Mg, Si, Ca, and Ti) are super solar and similar to those of normal halo stars. The [$\alpha$/Fe] ratio of our program stars lies between 0.35 and 0.43 dex. It is very close to normal halo stars that show an alpha enhancement of 0.4 dex. The abundances of odd-Z elements (Na and Al) fall in the range of other $r$-process enhanced stars. However, Na shows slight super solar abundance and Al displays little sub solar abundance. The iron peak elements (Sc, V, Cr, Mn, Fe, Co, Ni, Cu, Zn) also show abundances close to other $r$-process enhanced stars reported in the literature.

\subsection{Formation of neutron-capture elements}
\begin{figure}
    \centering
    \includegraphics[width=0.48\textwidth]{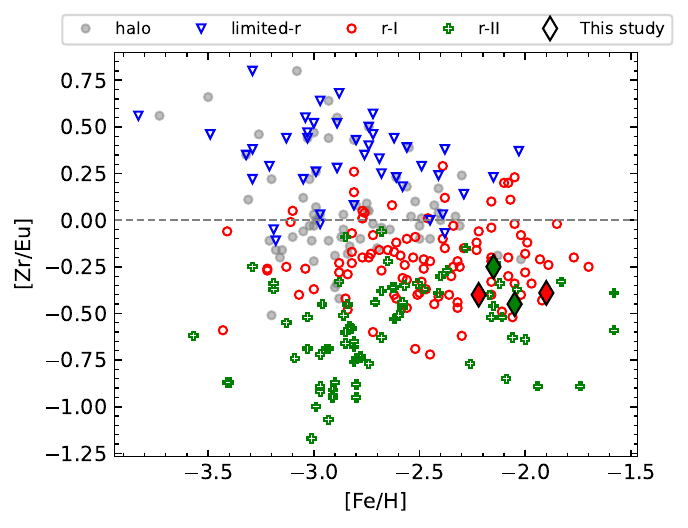}
    \includegraphics[width=0.48\textwidth]{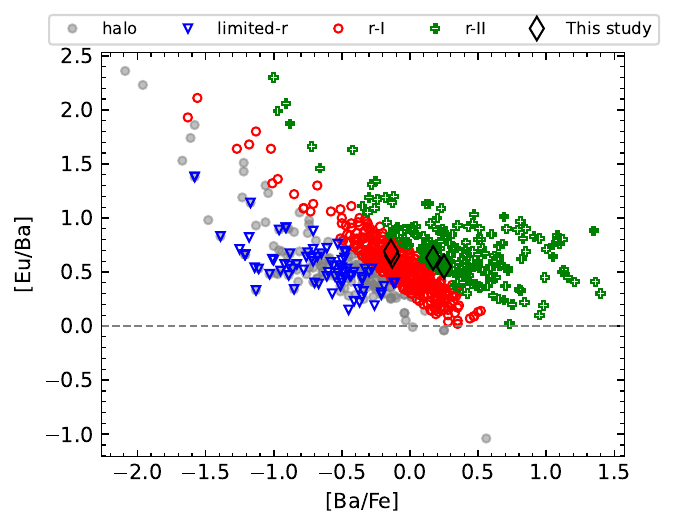}
    \caption{Left: [Zr/Eu] ratio as a function of [Fe/H]. Right: [Eu/Ba] ratio as a function of [Ba/Fe]. Red open circles show $r$-I stars, green open crosses display $r$-II stars, blue down triangles represent limited-$r$ stars and gray filled circles are normal halo stars. Big red and green faced diamonds are the stars from this study.}
    \label{fig:Fe_H__Zr_Eu_and_Ba_Fe__Eu_Ba}
\end{figure}

The elemental abundances of $r$-process enhanced stars as a function of atomic number or atomic mass show a unique pattern with three pronounced peaks that is commonly known as $r$-process pattern \citep{sneden.etal.2008}. To understand the production of the first $r$-process peak elements, we have plotted the evolution of [Zr/Eu] ratio as a function of metallicity in the left panel of figure~\ref{fig:Fe_H__Zr_Eu_and_Ba_Fe__Eu_Ba}. It shows high dispersion at low metallicity range, confirming the multiple channels of first peak element production. At low metallicity, we can also notice two fade branches which seem to join towards higher metallicity, further supporting the multiple channels of the first $r$-process peak nucleosynthesis. A large sample of $r$-process enhanced stars can help us constrain their formation channels.

In the right panel of figure~\ref{fig:Fe_H__Zr_Eu_and_Ba_Fe__Eu_Ba}, we have presented the evolution of [Ba/Eu] with [Ba/Fe] to investigate the origin of the second $r$-process peak elements. There can be seen a decreasing trend of [Eu/Ba] as we move towards a higher [Ba/Fe] value and then it becomes constant after [Ba/Fe] $=$ -0.5 dex. Also, a large fraction of $r$-process enhanced stars show [Ba/Fe] $>$ -0.5 dex. It means that Ba and Eu for the majority of $r$-process enhanced stars have originated in the same astrophysical conditions. At smaller [Ba/Fe], the sources of Ba and Eu are different for a small sample of $r$-process stars.

\begin{figure}
    \centering
    \includegraphics[width=\textwidth]{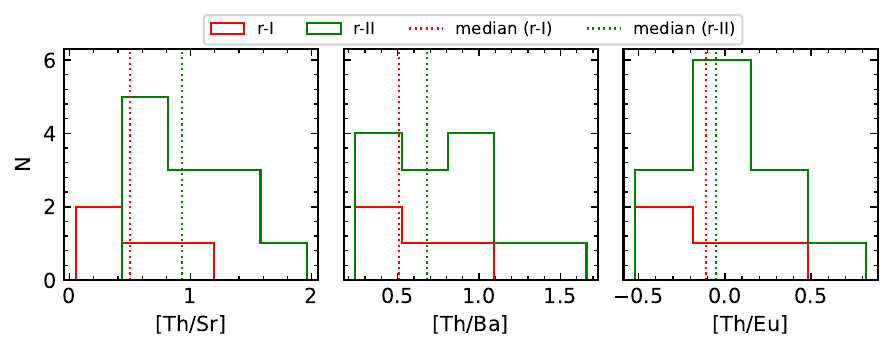}
    \caption{Distribution of [Th/Sr], [Th/Ba], and [Th/Eu] ratios in Th-detected $r$-process enhanced stars. The red histogram shows $r$-I stars, the green histogram displays $r$-II stars, and the dotted vertical lines are medians of corresponding distribution.}
    \label{fig:Th_distribution}
\end{figure}

Furthermore, we have calculated the distributions of [Th/Sr], [Th/Ba], and [Th/Eu] ratios for Sr-, Ba-, and Th-detected stars from literature including two objects from this study. Figure~\ref{fig:Th_distribution} shows these distributions. We found that the $r$-I stars show a lower median [Th/Sr] ratio than the $r$-II stars. The same behaviour is also noticed for the [Th/Ba] ratio but with a decreased difference. In the case of [Th/Eu], this difference is even much smaller. These results demonstrate the non-universal production of neutron-capture elements in two $r$-process sub-classes.

\subsection{Orbital kinematics}
To further investigate, we have calculated the stellar orbits of our program stars in the Milky Way potential using the publicly available code gala \citep{gala_code.2017}. For this purpose, we obtained the proper motions and parallaxes from Gaia DR3 \citep{Gaia_DR3.2022}.  Our analysis shows that out of four program stars, two (HD 263815 and TYC 1191-918-1) are falling in the disk region, and the rest two (TYC 3431-689-1 and TYC 1716-1548-1) are in the halo region. We found that TYC 1191-918-1 is on prograde orbit and the other three stars are on retrograde orbits. HD 263815 is part of the disk and is orbiting in the opposite direction of the Galactic rotation. This star might have accreted or formed in the halo and later migrated to the disk hence retaining the retrograde motion. In four of our program stars, one (TYC 3431-689-1) shows clear kinematics of ex-situ formation, and three are in mixed-zone according to \cite{Di_Matteo.etal.2020}. In Table~\ref{tab:kinematic_param}, we have listed the orbital parameters of our program stars.

\begin{table}[]
    \centering
    \begin{tabular}{|lcccccr|} \hline
         Object Name & $l_{xy}$ & $l_{z}$ & $r_{p}$ & $r_{a}$ & $z_{max}$ & $e$\\
          & (kpc km s$^{-1}$) & (kpc km s$^{-1}$) & (kpc) & (kpc) & (kpc)  & \\ \hline
        HD 263815 & 209.57 & -541.79 & 1.26 & 10.74 & 0.74 & 0.79 \\
        TYC 1191-918-1 & 240.66 & 482.23 & 1.15 & 10.36 & 0.94 & 0.80 \\
        TYC 3431-689-1 & 1506.37 & -1130.35 & 7.12 & 10.73 & 7.02 & 0.20\\
        TYC 1716-1548-1 & 1012.85 & -523.69 & 3.45 & 9.49 & 5.74 & 0.46 \\ \hline
    \end{tabular}
    \caption{Orbital parameters of our program stars obtained using gala code. Parameters $l_{x}$, $l_{y}$, and $l_{z}$ are components of specific angular momenta, $r_{p}$ is pericenter distance, $r_{a}$ is apocenter distance, $z_{max}$ is the maximum vertical height of the orbit, and $e$ is the eccentricity of the orbit.}
    \label{tab:kinematic_param}
\end{table}

\section{Conclusions}
\label{sec:conclusions}
We have presented the detailed elemental abundance analysis of four $r$-process enhanced stars and their kinematics. We obtained high-resolution spectra from HORuS on GTC. Thanks to the good signal-to-noise ratio, we could detect 16 light elements and 20 neutron-capture elements among four of our program stars. According to \cite{Holmbeck.etal.2018}, HD 263815 and TYC 1191-918-1 are $r$-I while TYC 3431-689-1 and TYC 1716-1548-1 are $r$-II. We found that our program stars show alpha enhancement similar to normal halo stars. Four of our program stars show [C/N] $<$ -1.0 dex and [N/Fe] $>$ +0.5 dex indicating the signature of mixing due to their evolution. The [Zr/Eu] ratio as a function of metallicity shows that the evidence of multiple channels for the production of $r$-process. Further, we showed the indications of multiple channels for Ba and Eu production at [Ba/Fe] $<$ -0.5 dex. Our results are tentative due to the limited sample towards lower metallicity and low [Ba/Fe] ratio. Th to Sr, Ba, and Eu ratio also support the similar non-universality of neutron-capture elements. An increased sample of $r$-process enhanced stars will enable us understand different formation channels of neutron capture elements. Using orbital analysis, we demonstrated that HD 263815 and TYC 1191-918-1 show disk-like kinematics, whereas TYC 3431-689-1 and TYC 1716-1548-1 show halo kinematics. HD 263815 is orbiting in the disk but on retrograde orbit, indicating its ex-situ origin. TYC 3431-689-1 displays clear ex-situ kinematics according to \cite{Di_Matteo.etal.2020} analogy.

\begin{acknowledgments}
PS and TS thank Carlos Allende Prieto for providing the data and Timothy C. Beers for helpful discussion during this project. This work uses Astropy \citep{astropy:2022}, Matplotlib \citep{Matplotlib.2007}, NumPy \citep{Numpy.2020}, Pandas \citep{Pandas.software.2020} and VOSA web-tool (https://svo.cab.inta-csic.es).
\end{acknowledgments}

\begin{furtherinformation}

\begin{orcids}
\orcid{0009-0001-4813-0432}{Pallavi}{Saraf}
\orcid{0000-0003-0891-8994}{Thirupathi}{Sivarani}


\end{orcids}

\begin{authorcontributions}
PS performed analysis, created visualizations, and wrote the draft under the supervision of TS.
\end{authorcontributions}

\begin{conflictsofinterest}
The authors declare no conflict of interest.
\end{conflictsofinterest}

\end{furtherinformation}

\bibliographystyle{bullsrsl-en}

\bibliography{extra}

\end{document}